\documentclass{elsart}
\usepackage{amsmath}
\journal{Astroparticle Physics}

\usepackage{graphicx}

\renewcommand{\it}{}

\begin{document}
\begin{frontmatter}
\title{Determination of the calorimetric energy in
extensive air showers}

\author{Henrique M.J. Barbosa\corauthref{cor}},
\corauth[cor]{Corresponding author.}
\ead{hbarbosa@ifi.unicamp.br}
\author{Fernando Catalani},
\author{Jos\'{e} A. Chinellato} and
\author{Carola Dobrigkeit}

\address{Departamento de Raios C\'osmicos e Cronologia,\\ Instituto de
F\'{\i}sica Gleb Wataghin, \\ Universidade Estadual de Campinas,\\ CP
6165 - 13083970, Campinas-SP, Brasil}

\begin{abstract}
The contribution of different components of an air shower to the total
energy deposit in the atmosphere, for different angles and primary
particles, was studied using the CORSIKA air shower simulation code.
The amount of missing energy, parameterized in terms of the
calorimetric energy, was calculated. The results show that this
parameterization varies less than $1\%$ with angle or observation
level.  The dependence with the primary mass is less than $5\%$ and,
with the high energy hadronic interaction model, less than $2\%$.
The systematic error introduced by the use of just one
parameterization of the missing energy correction function, for an
equal mixture of proton and iron at $45^o$, was calculated to be below
$3\%$. We estimate the statistical error due to shower-to-shower
fluctuations to be about $1\%$.
\end{abstract}
\end{frontmatter}

\section{Introduction}

In the last years, the quest to unravel the many mysteries related to
the cosmic radiation has been intensified, in an attempt to answer
open questions\,\cite{bib:rev_olinto,bib:rev_gaisser} about origin,
propagation and chemical composition of the ultra high energy cosmic
rays.  The Auger\,\cite{bib:tdr} observatory, now under construction,
and future experiments like Telescope Array\,\cite{bib:telarray},
EUSO\,\cite{bib:euso-artigo} and OWL\,\cite{bib:owl-artigo} will shed
light on these questions.

A common feature of these experiments is the use of the
fluorescence technique, first explored by the Fly's Eye
group\,\cite{bib:flyseye}. The fluorescence telescopes use
the atmosphere as a calorimeter, making a direct measurement of
the longitudinal shower development, which represents the most
appropriate technique to determine the energy of the primary
particle. This is based on the assumption that the fluorescence
yield is proportional to the local energy deposit. Recent
measurements\,\cite{bib:kakimoto,bib:nagano} have shown that this
is valid for electrons and efforts are being made to
improve\,\cite{bib:airfly,bib:flash,bib:aircamp} and extend the
results.

However, a fraction of the primary energy cannot be detected because
it is carried away by neutrinos and high energy muons that hit the
ground.  Corrections for the so called missing energy must be
properly applied to the calorimetric energy $E_{cal}$ measured, in
order to find the primary energy $E_0$. The first parameterizations of
the missing energy, as a function of the calorimetric energy, was done
by J.Linsley\,\cite{bib:linsley_edep0,bib:linsley_edep2} and the Fly's
Eye group\,\cite{bib:flyseye_edep}. Some years ago, C.\,Song et
al\,\cite{bib:song} recalculated it using simulated air showers.  The
method consisted of first determining the number of charged particles
in the shower $N_{ch}(t)$, {\it as a function of the atmospheric depth
$t$,} including a correction to take into account particles discarded
below the simulation threshold.  The track length integral was
then evaluated, assuming a mean energy loss rate
$\langle\alpha\rangle$, to find an estimate of the total calorimetric
energy:

\[
E_{cal} \simeq \langle\alpha\rangle \int_{0}^{\infty} N_{ch}(t) dt
\]

Nowadays, a very detailed simulation of the energy deposited in
the atmosphere by air showers\,\cite{bib:risse_dedx} is available
in the Monte Carlo program CORSIKA\,\cite{bib:corsika}.  In this
paper, we use it to study the contribution of different components
of an air shower to the total energy deposit. This approach
is different from the previous one\,\cite{bib:song} since we use
directly the energy deposited in the atmosphere, by each component
of the shower.

With this information, we calculate the amount of missing energy
and parameterize the fraction $E_{cal}/E_{0}$ as a function of
$E_{cal}$.  We derive this function for proton and iron primaries
and compare it with previous results. Additionally, we investigate
the dependence of this parameterization with the zenith angle,
high energy hadronic interaction model and observation level.

The outline of this paper is as follows: section \ref{sec:simul} gives
a brief description of the simulations performed.  In section
\ref{sec:emiss} we discuss the concepts of missing energy and energy
deposit as ingredients of our calculations to obtain the primary
energy. In section \ref{sec:results}, we present a summary of the
longitudinal profiles of energy deposit generated in the simulations
and we discuss the missing energy correction curve obtained. Our final
conclusions are presented in section \ref{sec:conclusions}.

\section{Simulation}
\label{sec:simul}

The CORSIKA simulation code, version 6.30, was used to study
the energy balance in the propagation of atmospheric showers. This
Monte Carlo package, designed to simulate extensive air showers up
to the highest energies, is able to track in great detail almost
all particles.

In particular, this program provides tables for the energy
dissipated in the atmosphere as a function of the atmospheric
depth\,\cite{bib:risse_dedx}. For each kind of particle (gamma,
electron, muon, hadron and neutrino), the energy lost by
ionization in air and by simulation cuts (angle and energy
thresholds) are specified. In addition, there is also information
about the amount of energy carried by particles hitting the
ground.

In this paper, QGSJET01\,\cite{bib:qgsjet,bib:crsk_qgsjet} has been
employed for higher energies $E_{lab}>80GeV$, and
GHEISHA\,\cite{bib:geisha} for lower energies.
Optimum\,\cite{bib:optthin1,bib:optthin2} thinning\,\cite{bib:thin}
was applied, so as to reduce CPU time, while keeping the simulation
detailed enough for our purposes. {\it We have used a thinning level
of $\varepsilon_{th} = 10^{-4}$, with maximum weight limitation set to
$\omega_{max} = 10^{-4}E_0/GeV$ for muons, hadrons and electromagnetic
particles.  The energy threshold for electromagnetic particles was set
to $E_{thr}^{em} = 50keV$ and, for hadrons and muons, $E_{thr}^{h} =
E_{thr}^\mu = 50MeV$.}  Ground was fixed at sea level.

Proton and iron initiated air showers were simulated for three
different energies $10^{18}eV$, $10^{19}eV$ and $10^{20}eV$ and four
different angles $0^o$, $30^o$, $45^o$ and $60^o$. {\it For each
combination of these parameters, $100$ showers were simulated and the
longitudinal profiles of energy deposit were carefully studied.}

{\it With the angle fixed at $45^o$, we have also simulated proton and
iron showers for two different conditions. First, with a different
high energy hadronic interaction model
(SIBYLL2.1\,\cite{bib:sibyll0,bib:sibyll1,bib:sibyll}). Second, with a
different detection level (300m above sea level)}.

\section{Energy Balance}
\label{sec:emiss}

We will use a nomenclature for energy distribution that is similar to
that usually seen in analytic solutions of diffusion equations.  The
number of particles of type $i$ at depth $t$, with energy in the
interval $[E, E + dE]$, is $N_i (E,t) dE$.  Therefore, $F_i (E,t) dE =
E N_i (E,t) dE$ is the total energy of these particles.  The integral
form of these distributions is written as $F_i (E < E_0, t ) dt =
\int_{0}^{E_0} F_i (E,t) dE$ and so on.

For the shower processes, we define $i = h$ (for hadrons), $\mu$ (for
muons), $\nu$ (for neutrinos), $em$ (electromagnetic component).  For
the $i_{th}$ component, the energy deposited due to ionization prior
to depth $t$ is given by $D_i(t)$.

Then, at any depth, we can write an expression for the primary
energy:

\[
E_0=\sum_{i=h,\mu,\nu,em}F_i(E>0,t) + \sum_{i=h,\mu,\nu,em}D_i(t)
\]

Now, if we introduce energy thresholds\footnote{Even though there
are also angular cuts, we write only $E \leq E_{thr}$, for
simplicity.} we have

\[
E_0=\sum_{i}F_i(E>E_{thr}^i,t) + 
\sum_{i}F_i(E\leq E_{thr}^i,t) + \sum_{i}D_i(t)
\]

For practical reasons, particles submitted to threshold cuts (both in
energy and in angle) are no longer tracked, and this expression is not
exact.  {\it CORSIKA computes the energy of these particles as
dissipated energy.  The exact equation must consider the sum of the
energies of all particles below the threshold $F_i^\ast(E\leq
E_{thr}^i,t)$, prior do depth $t$, and the energy deposited only by
particles above the threshold $D_i^\ast(t)$:}

\[
E_0=\sum_{i} F_i (E>E_{thr}^i,t)
+ \sum_{i} F_i^\ast(E\leq E_{thr}^i,t) + \sum_{i} D_i^\ast(t)
\]

Now we can define calorimetric and missing energy for our purposes.
We take as calorimetric energy the energy deposited in air by
ionization. The amount of energy which does not cause ionization is
the missing energy. {\it The energy in the electromagnetic component
and a fraction $f^\prime_h$ of the energy in the hadronic component,
no matter the stage of shower development, are also taken as
calorimetric.  This has to be so, because they are accounted for when
evaluating the track length integral.}

{\it Besides that, when a particle reaches a threshold, its energy is
dropped from the simulation. However, a fraction of this energy would
in reality go to ionization.} We can take these fractions as
constants, for each component, as they are averaged over a large
number of low energy particles.  {\it Therefore, $f_i F_i^\ast(E\leq
E_{thr}^{i},t)$ is the fraction of the sub-threshold particles
energy which goes to ionization.} The remaining ($1-f_i$) goes to
missing energy. {\it The values of the constants $f_i$ and $f^\prime_h$
will be given later.}

From the above equations, we can define the calorimetric energy
at depth $t$ as:

\begin{eqnarray*}
E_{cal}(t) & = & F_{em} (E>E_{thr},t)+F_{em}^\ast(E\leq E_{thr},t) + \\
           &   & f^\prime_h F_h(E>E_{thr}^h,t) + f_h F_h^\ast(E\leq E_{thr}^{h},t) + \\
           &   & f_\mu F_\mu^\ast(E\leq E_{thr}^{\mu},t) + \sum_i D_i^\ast (t)
\end{eqnarray*}

The remaining terms will add to the missing energy:

\begin{eqnarray*}
E_{miss}(t)
& = & (1-f^\prime_h) F_h ( E>E_{thr}^h,t) + (1-f_h) F_h^\ast(E\leq E_{thr}^{h},t) + \\
&   &  F_\mu (E>E_{thr}^{\mu},t)+(1-f_{\mu})F_\mu^\ast(E\leq E_{thr}^{\mu},t) + \\
&   &  F_\nu(E>0,t)
\end{eqnarray*}

We will consider now the fractions going to calorimetric and
missing energies for particles submitted to threshold cuts. 
It is important to know the type of particles contributing and
since this information was not available, we modified the CORSIKA
code. {\it For each particle dropped from the simulation, its type,
kinetic energy and releasable energy were recorded. The releasable 
energy is the energy that would have been lost had the particle
not been dropped from the simulation.

Thus we could calculate the contributions of the different types of
particles and their respective energy spectrum. Table~\ref{tab:aircut}
shows the mean kinetic energy and the relative contributions
calculated from 500 proton showers with $10^{19}eV$ and $45^o$. The
relative contribution is the ratio of the energy deposited by a given
type of particle to the total energy deposited by the respective
shower component.} The thinning level was set to $10^{-4}$
without weight limitation. We found similar results for iron induced
air shower simulations.

{\it This information was used to further track these sub-threshold
particles using the Geant4 code~\cite{bib:geant}.} For each type of
particle, 5000 air shower simulations initiated by these low energy
particles were performed.  The energy spectrum obtained from CORSIKA
was used to set the energy of these events and the secondary particles
were followed through a sea level standard atmosphere until
$1keV$. The Geant package allowed us to find out the amount of energy
lost by ionization and other processes, and determine the fractions of
energy deposited by ionization. {\it These fractions are also shown in 
table\,\ref{tab:aircut}}. For the low energy
electromagnetic particles, we made the assumption that all the
energy is deposited ($f_{em}=1$). For muons and hadrons, we used
the fractions $f_\mu=0.425$ and $f_h=0.739$.

{\it It is also necessary to consider what happens to the hadrons
hitting the ground. They have a significantly different composition
and energy spectrum from those hadrons cut in air and we cannot assume
that $f^\prime_h = f_h$.  We used the CORSIKA information about
particles at detection level to analyze the energy spectrum and
relative contributions of different types of hadrons.
Table~\ref{tab:grdcut} shows the results for the same set of 500 CORSIKA
showers. In this case, the average energies are higher and we found
the dependence of $f^\prime_h$ on energy to be a second order
effect. Geant4 was then used to simulate 500 showers, for each
particle, with energy fixed to its average kinetic energy.}  We found
that $f^\prime_h=0.61$.

{\it Therefore, the prescription used for calculating
the missing energy at ground level is:}

\begin{eqnarray}
E_{miss}
= 0.390 F_h( E>E_{thr}^h,grd)&+&0.261 F_h^\ast(E\leq E_{thr}^{h}, grd) + \nonumber \\
  1.000 F_\mu(E>E_{thr}^{\mu},grd)&+&0.575 F_\mu^\ast(E\leq E_{thr}^{\mu}, grd) + 
\label{eq:emiss} \\ & & 1.000 F_\nu(E>0,grd) \nonumber
\end{eqnarray}

Since $E_{cal}/E_0 = 1 - E_{miss}/E_0$, we can calculate the
fraction of the primary energy measured with the fluorescence
telescopes.


\section{Results}
\label{sec:results}

For the sake of clarity, we present in this section the results from
our simulations.  In table \ref{tab:results}, we show the contribution
for the energy deposit from different shower components, as a
percentage of the primary energy, for each simulated angle and energy.

Each three lines, corresponding to a fixed choice of angle and energy,
refer to three different contributions to the energy deposit, which we
called: (ION) ionization in air, (CUT) simulation cuts, and (GRD)
particles arriving at ground. Each column gives the contribution from
different shower components: gammas, electrons, muons, hadrons and
neutrinos. The first value (first/second) refers to proton primaries,
while the second refers to iron primaries.

The first line accounts for ionization energy losses in air,
consequently we have no contributions from gammas or neutrinos. We
considered all contributions from this line as calorimetric energy.

The second line gives the total energy of particles left out of the
simulations because of angular or energy cuts. A fraction of this
energy will go to ionization, as described in last section. Since
neutrinos are not tracked in the simulation, they are dropped as soon
as produced and all their energy appears in this line. We considered
all their energy as missing energy.

The last line gives the amount of energy carried by particles arriving
at the observation level.  In this case, we considered all energy from
the electromagnetic and a fraction of the hadronic component hitting
the ground as calorimetric energy. The energy carried by muons was
taken as missing energy.

\subsection{Missing energy}

We have used the data in table\,\ref{tab:results} to evaluate the mean
missing energy for each combination of angle, energy and primary
particle as given by equation \ref{eq:emiss}.

\begin{figure}[p]
\centerline{\includegraphics[clip=true,angle=0,width=5in]{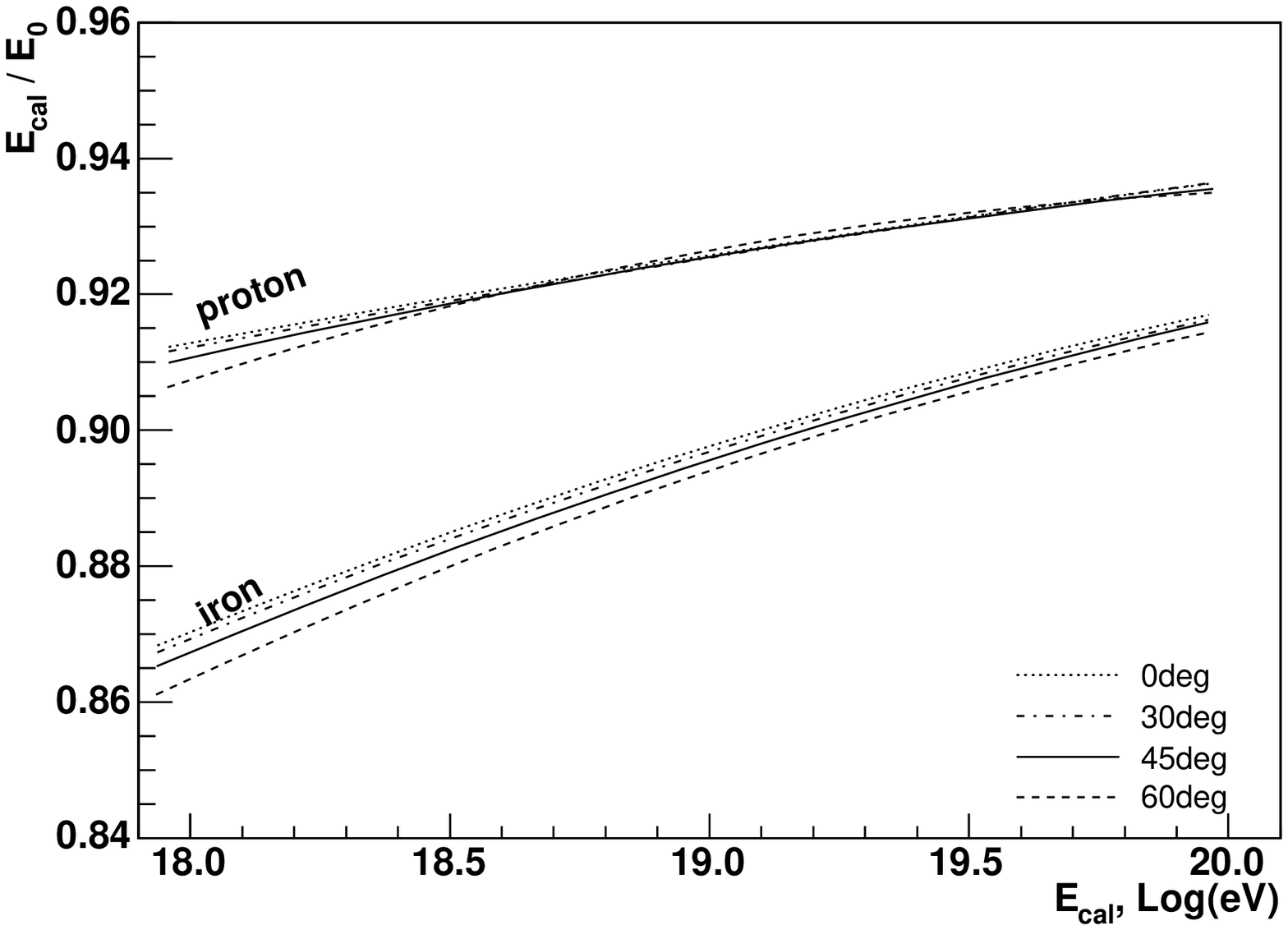}}
\caption{Missing energy correction plotted as the
fraction $E_{cal}/E_0$ as a function of $E_{cal}$. The curves are
for proton and iron showers at four different angles.}
\label{fig:emiss23}
\end{figure}

In figure \ref{fig:emiss23}, the correction factor for the missing
energy is shown, plotted as the fraction $E_{cal}/E_0$, as a function
of $E_{cal}$. We can see that the variation in the amount of missing
energy is only slightly dependent on the angle and decreases with
energy. At $10^{18}eV$, the ratio $E_{cal}/E_0$ varies between
$90.8-91.3\%$ for protons, and between $86.4-87.0\%$ for iron.

On the other hand, the ratio $E_{cal}/E_0$ is largely dependent on the
primary mass. At a fixed angle, the difference in the amount of
missing energy for proton and iron showers decreases with energy. For
instance, at $45^o$ this difference\footnote{In this and in the next
section, all differences are relative, e.g., $(v_1 - v_2) / \bar{v}$.}
is $4.9\%$ at $10^{18}eV$, $3.3\%$ at $10^{19}eV$ and $2.1\%$ at
$10^{20}eV$.

A mean parameterization of $E_{cal}/E_0$, taken as a mixture
of $50\%$ proton and $50\%$ iron at $45^o$, is usually used in the
reconstruction routines. In table \ref{tab:emiss_fit}, we give
the fitting parameters of the mean missing energy correction
function for different simulations conditions. For showers with
zenith angle $45^o$, we have:

\begin{equation}
\frac{E_{cal}}{E_0} = 0.967 - 0.078 \left(\frac{E_{cal}}{1EeV}\right)^{-0.140}
\end{equation}

The systematic error in the value of $E_{cal}/E_0$, when
calculating the primary energy with the mean parameterization, is less
than $\pm2.5\%\pm0.4\%$ at $1EeV$ and decreases with energy.
Shower-to-shower fluctuations were evaluated, and found to be
independent of energy. The {\it rms} value of $E_{cal}$ is $1.1\%$ of
the primary energy for proton showers, and $0.4\%$ for iron showers.

\begin{figure}[p]
\centerline{\includegraphics[clip=true,angle=0,width=5in]{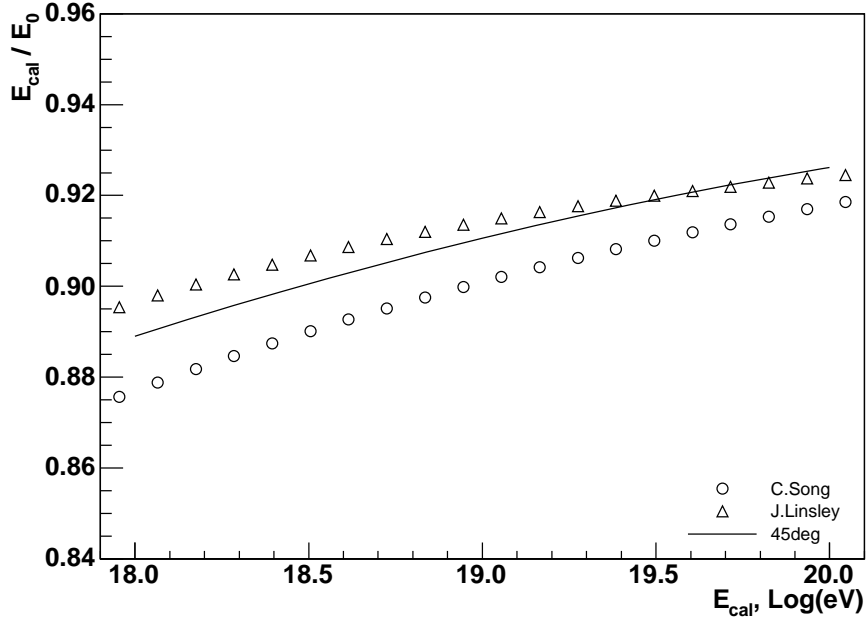}}
\caption{Missing energy correction plotted as the fraction
$E_{cal}/E_0$ as a function of $E_{cal}$.  The average between iron and
proton missing energy correction is plotted. For comparison,
Song's and Linsley's results are also shown. }
\label{fig:emissmean}
\end{figure}

\begin{figure}[p]
\centerline{\includegraphics[clip=true,angle=0,width=5in]{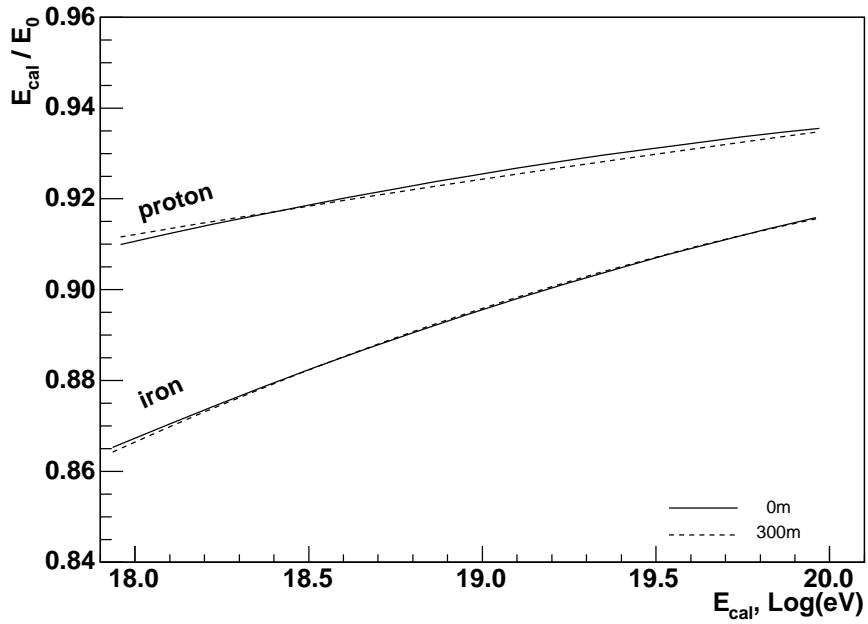}}
\caption{Missing energy correction plotted as the fraction
$E_{cal}/E_0$ as a function of $E_{cal}$. The dependence of the
parameterization on ground level on ground level is shown.
Simulations for proton and iron primaries, at $45^o$.}
\label{fig:emiss_diff1}
\end{figure}

Figure \ref{fig:emissmean} shows in detail the mean missing energy
correction, calculated for a mixture of 50\% proton and 50\% iron.  As
can be noticed, there is a small shift in the curves relative to the
previous results by Song\,\cite{bib:song} and
Linsley\,\cite{bib:linsley_edep2}. Song's results are based on
Monte Carlo simulations, and the mixed composition is shown.
Linsley's results were calculated based on muon size measurements, and
are independent of the primary composition. Results agree to within
$1\%$, but they were not obtained under the same conditions. In
figure \ref{fig:emiss_diff1}, we compare, for $45^o$, the missing
energy correction curves obtained for the ground level set to $0m$ and
$300m$, which is the same used by Song et al. As can be seen, our
result does not depend on the observation level.

\begin{figure}[p]
\centerline{\includegraphics[clip=true,angle=0,width=5in]{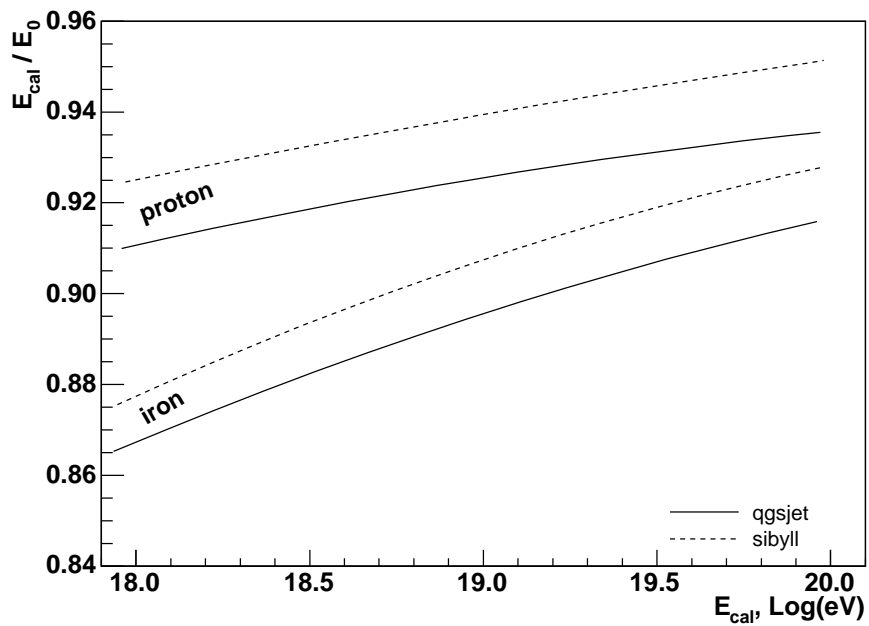}}
\caption{Missing energy correction plotted as the fraction
$E_{cal}/E_0$ as a function of $E_{cal}$. The variation with the
high energy hadronic interaction model is shown. Simulations for
proton and iron primaries, at $45^o$. } \label{fig:emiss_diff2}
\end{figure}

The dependence on the interaction model is considerable, as shown in
figure \ref{fig:emiss_diff2}.  We compared, for proton and iron at
$45^o$, the missing energy correction curves obtained with QGSJET01
and SIBYLL2.1 models. The difference is practically constant: around
$1.6\%$ for proton and $1.2\%$ for iron. In all cases, SIBYLL predicts
less missing energy than QGSJET.

\section{Conclusions}
\label{sec:conclusions}

The energy deposit by atmospheric air showers was studied
aiming for a better reconstruction of the primary energy. The new and
very detailed energy balance now present in CORSIKA was used.

Our results for the missing energy correction function agree with
previous calculations\,\cite{bib:linsley_edep2,bib:song}, to within
$1\%$. The dependence of this function on angle and observation level
was found to be less than $0.7\%$. Comparing the high energy hadronic
interaction models QGSJET and SIBYLL, the results differ by less than
$1.6\%$.  The larger dependence comes from the primary mass, being less
than $5\%$ at $10^{18}eV$, decreasing with energy.

We found a mean parameterization of $E_{cal}/E_0$, taken as a
mixture of $50\%$ proton and $50\%$ iron at $45^o$, as usually
used in the reconstruction routines.  Considering the QGSJET model
only, we estimate that the total systematic error, introduced by
the use of just this parameterization, is below $3\%$ at $1EeV$,
and below $2\%$ at $100EeV$. This is of the order of
shower-to-shower fluctuations. For proton showers, the {\it rms}
value of $E_{cal}$ is $1.1\%$ of the primary energy. For iron
showers, it is $0.4\%$.

\ack

The authors would like to thank Bruce Dawson, Chihwa Song, Dieter Heck
and Markus Risse for the fruitful discussions.  This work was
supported by the Brazilian science foundations CNPq and CAPES to which
we are grateful. The calculations were done using computational
facilities in Campinas funded by FAPESP.

\begin{table}[p]
\begin{center}
\begin{tabular}{ccccc}
\hline\hline
particle  &  $E_{kin}\ (MeV)$ & Rel. Contri. (\%) & Ion. Frac. (\%) &\\
\hline
$\gamma$  &  .458             & 21.0 &  99.7 $\pm$ 0.4 &\\
$e^+$     &  4.29             & 20.9 &  99.7 $\pm$ 0.3 &$f_{em}=99.8$\\
$e^-$     &  1.87             & 58.1 &  99.8 $\pm$ 0.5 &\\
\hline
$\mu^+$   &  8.30             & 48.7 &  43.  $\pm$ 14. &\\
$\mu^-$   &  8.26             & 51.3 &  42.  $\pm$ 14. &\raisebox{1.5ex}[0pt]{$f_\mu=42.5$} \\
\hline
$n$       &  43.2             & 22.9 &  57.  $\pm$ 38. &\\
$p$       &  35.6             & 18.0 &  98.  $\pm$  9. &\\
$\pi^0$   &  86.0             & 27.5 &  99.7 $\pm$ 0.1 &\\
$\pi^-$   &  88.6             & 14.6 &  45.  $\pm$ 16. &\\
$\pi^+$   &  95.6             & 13.7 &  47.  $\pm$ 16. &\raisebox{1.5ex}[0pt]{$f_h=73.9$}\\
$^2H$     &  37.2             &  1.7 &  99.  $\pm$ 5.  &\\
$^3H$     &  43.2             &  0.8 &  99.  $\pm$ 6.  &\\
$^4H$     &  41.4             &  0.3 &  99.  $\pm$ 3.  &\\
\hline\hline
\end{tabular}
\caption{Particles discarded in air. Mean kinetic energy, relative
contribution and fraction of energy going to ionization are
shown. The fractions $f_{em}$, $f_\mu$ and $f_h$ are the average
over all particles weighted by their relative contributions.}
\label{tab:aircut}
\end{center}
\end{table}

\begin{table}[p]
\begin{center}
\begin{tabular}{ccccc}
\hline\hline
particle   &  $E_{kin}\ (GeV)$ & Rel. Contri. (\%) & Ion. Frac(\%) &\\
\hline
$n$        &  10.              &   7.12   &   70.1 $\pm$ 0.1  &\\
$p$        &  31.6             &   5.57   &   75.3 $\pm$ 0.1  &\\
$\bar{p}$  &  100.             &   3.21   &   73.2 $\pm$ 0.1  &\\
$\bar{n}$  &  100.             &   2.92   &   72.2 $\pm$ 0.1  &\\
$K^0_L$    &  1000.            &   7.30   &   36.3 $\pm$ 0.2  &\raisebox{1.5ex}[0pt]{$f^\prime_h=61.0$}\\
$K^\pm$    &  1000.            &   8.86   &   60.4 $\pm$ 0.2  &\\
$\pi^-$    &  316.             &  32.06   &   61.7 $\pm$ 0.2  &\\
$\pi^+$    &  316.             &  33.01   &   59.3 $\pm$ 0.2  &\\
\hline\hline
\end{tabular}
\caption{Hadrons at ground level. Mean kinetic energy, relative
contribution and fraction of energy going to ionization are
shown. The fraction $f^\prime_h$ is the average over all particles
weighted by their relative contributions.} \label{tab:grdcut}
\end{center}
\end{table}

\renewcommand{\baselinestretch}{0.75}\large\normalsize
\begin{table}[p]
\begin{center}
\begin{tabular}{r|c|c|ccccc}
\multicolumn{3}{c}{ } & gammas       & electrons        & muons            & hadrons          & neutrinos\\
\hline
        &        & ION  &   -  /  -   & 64.7 / 65.3 &  1.0 /  1.5 &  0.2 /  0.3 &   -  /  -        \\
        &1EeV    & CUT  &  1.1 /  1.2 & 10.7 / 10.8 &  0.1 /  0.1 &  0.3 /  0.5 &  3.0 /  4.6 \\     
        &        & GRD  &  8.4 /  4.6 &  4.2 /  2.1 &  5.2 /  8.0 &  1.1 /  1.1 &   -  /  -        \\
\cline{2-8}		                                                                             
        &        & ION  &   -  /  -   & 62.0 / 64.9 &  0.8 /  1.2 &  0.2 /  0.3 &   -  /  -        \\
$0^o$   &10EeV   & CUT  &  1.1 /  1.2 & 10.2 / 10.7 &  0.1 /  0.1 &  0.3 /  0.4 &  2.5 /  3.6 \\     
        &        & GRD  & 11.3 /  7.0 &  5.9 /  3.4 &  4.4 /  6.2 &  1.1 /  1.2 &   -  /  -        \\
\cline{2-8}		                                                                             
        &        & ION  &   -  /  -   & 57.7 / 62.8 &  0.7 /  1.0 &  0.2 /  0.2 &   -  /  -        \\
        &100EeV  & CUT  &  1.0 /  1.1 &  9.4 / 10.3 &  0.1 /  0.1 &  0.3 /  0.3 &  2.1 /  2.8 \\     
        &        & GRD  & 15.2 / 10.1 &  8.4 /  5.1 &  3.7 /  4.9 &  1.2 /  1.3 &   -  /  -        \\
\hline			                                                                             
\hline			                                                                             
        &        & ION  &   -  /  -   & 69.8 / 67.6 &  1.2 /  1.8 &  0.2 /  0.3 &   -  /  -        \\
        &1EeV    & CUT  &  1.3 /  1.3 & 13.0 / 12.7 &  0.1 /  0.1 &  0.4 /  0.5 &  3.3 /  5.0 \\     
        &        & GRD  &  3.4 /  1.7 &  1.5 /  0.7 &  5.3 /  8.0 &  0.5 /  0.4 &   -  /  -        \\
\cline{2-8}		                                                                             
        &        & ION  &   -  /  -   & 69.0 / 69.1 &  1.0 /  1.4 &  0.2 /  0.3 &   -  /  -        \\
$30^o$  &10EeV   & CUT  &  1.3 /  1.3 & 12.9 / 12.9 &  0.1 /  0.1 &  0.4 /  0.4 &  2.7 /  3.9 \\     
        &        & GRD  &  5.0 /  2.7 &  2.4 /  1.2 &  4.4 /  6.2 &  0.5 /  0.5 &   -  /  -        \\
\cline{2-8}		                                                                             
        &        & ION  &   -  /  -   & 67.6 / 69.5 &  0.8 /  1.1 &  0.2 /  0.2 &   -  /  -        \\
        &100EeV  & CUT  &  1.3 /  1.3 & 12.6 / 13.0 &  0.1 /  0.1 &  0.3 /  0.4 &  2.3 /  3.1 \\     
        &        & GRD  &  7.0 /  4.0 &  3.6 /  1.8 &  3.7 /  5.0 &  0.5 /  0.5 &   -  /  -        \\
\hline			                                                                             
\hline			                                                                             
        &        & ION  &   -  /  -   & 71.2 / 67.2 &  1.4 /  2.0 &  0.2 /  0.3 &   -  /  -        \\
        &1EeV    & CUT  &  1.5 /  1.4 & 15.6 / 14.7 &  0.1 /  0.1 &  0.4 /  0.5 &  3.6 /  5.4 \\     
        &        & GRD  &  0.5 /  0.3 &  0.2 /  0.1 &  5.2 /  7.9 &  0.1 /  0.1 &   -  /  -        \\
\cline{2-8}		                                                                             
        &        & ION  &   -  /  -   & 72.3 / 69.8 &  1.2 /  1.6 &  0.2 /  0.2 &   -  /  -        \\
$45^0$  &10EeV   & CUT  &  1.5 /  1.5 & 15.8 / 15.3 &  0.1 /  0.1 &  0.4 /  0.5 &  3.0 /  4.2 \\     
        &        & GRD  &  0.8 /  0.4 &  0.3 /  0.2 &  4.3 /  6.1 &  0.1 /  0.1 &   -  /  -        \\
\cline{2-8}		                                                                             
        &        & ION  &   -  /  -   & 72.7 / 71.5 &  1.0 /  1.3 &  0.2 /  0.2 &   -  /  -        \\
        &100EeV  & CUT  &  1.5 /  1.5 & 15.9 / 15.7 &  0.1 /  0.1 &  0.4 /  0.4 &  2.6 /  3.4 \\     
        &        & GRD  &  1.2 /  0.7 &  0.5 /  0.3 &  3.7 /  4.8 &  0.1 /  0.1 &   -  /  -        \\
\hline			                                                                             
\hline			                                                                             
        &        & ION  &   -  /  -   & 67.5 / 63.4 &  1.7 /  2.3 &  0.2 /  0.2 &   -  /  -        \\
        &1EeV    & CUT  &  1.7 /  1.6 & 19.2 / 18.1 &  0.1 /  0.1 &  0.5 /  0.6 &  4.1 /  6.1 \\     
        &        & GRD  &  0.0 /  0.0 &  0.0 /  0.0 &  5.1 /  7.6 &  0.0 /  0.0 &   -  /  -        \\
\cline{2-8}		                                                                             
        &        & ION  &   -  /  -   & 69.3 / 66.2 &  1.4 /  1.9 &  0.1 /  0.2 &   -  /  -        \\
$60^0$  &10EeV   & CUT  &  1.7 /  1.7 & 19.7 / 18.8 &  0.1 /  0.1 &  0.4 /  0.5 &  3.3 /  4.7 \\     
        &        & GRD  &  0.0 /  0.0 &  0.0 /  0.0 &  4.0 /  5.8 &  0.0 /  0.0 &   -  /  -        \\
\cline{2-8}		                                                                             
        &        & ION  &   -  /  -   & 70.1 / 68.2 &  1.2 /  1.6 &  0.1 /  0.2 &   -  /  -        \\
        &100EeV  & CUT  &  1.8 /  1.7 & 19.9 / 19.4 &  0.1 /  0.1 &  0.4 /  0.5 &  2.9 /  3.8 \\     
        &        & GRD  &  0.0 /  0.0 &  0.0 /  0.0 &  3.5 /  4.6 &  0.0 /  0.0 &   -  /  -        \\
\hline
\end{tabular}
\caption{Mean energy deposit contributions (in percentage of primary's
energy) from different shower components.  Each three lines correspond
to a fixed choice of angle and energy, and discriminate: (ION)
ionization in air, (CUT) simulation cuts and (GRD) particles arriving
at ground.  The first value (first/second) refers to proton showers while the
second refers to iron showers.}
\label{tab:results}
\end{center}
\end{table}
\renewcommand{\baselinestretch}{1.0}\large\normalsize

\begin{table}[p]
\begin{center}
\begin{tabular}{r|ccc|ccc|ccc}
\multicolumn{10}{l}{Coefficients of correction function} \\ \hline
\multicolumn{1}{c}{} & \multicolumn{3}{c}{Iron}&
\multicolumn{3}{c}{Iron/Proton}& \multicolumn{3}{c}{Proton}\\ \hline
angle &   A   &   B   &   C   &   A   &   B   &   C   &   A   &   B   &   C   \\ \hline
 0  &  0.970 & 0.100 & 0.139 & 0.977 & 0.085 & 0.117 & 0.984 & 0.071 & 0.089 \\
30  &  0.971 & 0.102 & 0.137 & 0.979 & 0.088 & 0.114 & 0.986 & 0.074 & 0.088 \\
45  &  0.977 & 0.109 & 0.130 & \bf 0.967 & \bf 0.078 & \bf 0.140 & 0.958 & 0.048 & 0.162 \\
60  &  0.962 & 0.098 & 0.161 & 0.948 & 0.062 & 0.220 & 0.942 & 0.035 & 0.337 \\
\hline
\end{tabular}
\caption{Fitting parameters for different simulation conditions, as
plotted in figure \ref{fig:emiss23}. The mid column indicates the
$50\%/50\%$ mixture and the values corresponding to $45^o$ are bold
faced. The fit function used was: $A - B(E/EeV)^{-C}$.}
\label{tab:emiss_fit}
\end{center}
\end{table}

\bibliographystyle{elsart-num}
\bibliography{barbosa_astphy951}

\begin{thebibliography}{10}
\expandafter\ifx\csname url\endcsname\relax
  \def\url#1{\texttt{#1}}\fi
\expandafter\ifx\csname urlprefix\endcsname\relax\def\urlprefix{URL }\fi

\bibitem{bib:rev_olinto}
A.~V. Olinto, Phys. Rep. 333-334 (2000) 329.

\bibitem{bib:rev_gaisser}
T.~K. Gaisser, Nucl. Phys. B (Proc. Suppl.) 117 (2003) 318.

\bibitem{bib:tdr}
\mbox{J. Cronin for the Auger Collaboration}, {T}echnical {D}esign {R}eport,
  The Pierre Auger Observatory, www.auger.org (1996).

\bibitem{bib:telarray}
\mbox{T. Yamamoto for the Telescope Array Collaboration}, Nucl. Instr. Meth. A
  488 (2002) 191.

\bibitem{bib:euso-artigo}
\mbox{M. Teshima for the EUSO Collaboration}, in: {P}roc. 28th {ICRC}, Tsukuba,
  2003, p. 1069.

\bibitem{bib:owl-artigo}
P.~Dierickx, et~al., in: {P}roc. B\"ackaskog Workshop on Extremely Large
  Telescopes, 2000, p.~43.

\bibitem{bib:flyseye}
R.~M. Baltrusaitis, et~al., Nucl. Instr. Meth. A 240 (1985) 410.

\bibitem{bib:kakimoto}
F.~Kakimoto, et~al., Nucl. Instr. Meth. A 372 (1996) 527.

\bibitem{bib:nagano}
M.~Nagano, K.~Kobayakawa, N.~Sakaki, K.~Ando, Astropart. Phys. 20 (2003) 293.

\bibitem{bib:airfly}
F.~Arciprete, et~al., in: {P}roc. 28th {ICRC}, Tsukuba, 2003, p. 837.

\bibitem{bib:flash}
\mbox{P. H\"untemeyer for the {FLASH} Collaboration}, in: {P}roc. 28th {ICRC},
  Tsukuba, 2003, p. 845.

\bibitem{bib:aircamp}
E.~Kemp, H.~Nogima, L.~G. dos Santos, C.~O. Escobar, A.~C. Fauth, in: {P}roc.
  28th {ICRC}, Tsukuba, 2003, p. 853.

\bibitem{bib:linsley_edep0}
J.~Linsley, in: {P}roc. 18th {ICRC}, Vol.~12, Bangalore, 1983, p. 135.

\bibitem{bib:linsley_edep2}
J.~Linsley, in: {P}roc. 19th {ICRC}, Vol.~2, La Jolla, 1985, p. 154.

\bibitem{bib:flyseye_edep}
R.~M. Baltrusaitis, et~al., in: {P}roc. 19th {ICRC}, Vol.~7, La Jolla, 1985, p.
  159.

\bibitem{bib:song}
C.~Song, Z.~Cao, B.~R. Dawson, B.~E. Fick, P.~Sokolsky, X.~Zhang, Astropart.
  Phys. 14 (2000) 7.

\bibitem{bib:risse_dedx}
M.~Risse, D.~Heck, Astropart. Phys. 20 (2004) 661.

\bibitem{bib:corsika}
D.~Heck, J.~Knapp, J.~Capdevielle, G.~Schatz, T.~Thouw, Tech. Rep. {FZKA} 6019,
  Forschungszentrum Karlsruhe, www-ik.fzk.de/$\sim$heck/corsika (1998).

\bibitem{bib:qgsjet}
N.~N. Kalmykov, S.~S. Ostapchenko, A.~I. Pavlov, Nucl. Phys. B (Proc. Suppl.)
  52B (1997) 17.

\bibitem{bib:crsk_qgsjet}
D.~Heck, et~al., in: {P}roc. 27th {ICRC}, Vol.~1, Hamburg, 2001, p. 233.

\bibitem{bib:geisha}
H.~Fesefeldt, Tech. Rep. {PITHA}-85/02, RWTH Aachen (1985).

\bibitem{bib:optthin1}
\mbox{M. Kobal for the Pierre Auger Collaboration}, Astropart. Phys. 15 (2001)
  259.

\bibitem{bib:optthin2}
M.~Risse, D.~Heck, J.~Knapp, S.~Ostapchenko, in: {P}roc. 27th {ICRC}, Vol.~2,
  Hamburg, 2001, p. 522.

\bibitem{bib:thin}
A.~M. Hillas, Nucl. Phys. B (Proc. Suppl.) 52B (1997) 29.

\bibitem{bib:sibyll0}
R.~Engel, T.~Gaisser, P.~Lipari, T.~Stanev, in: {P}roc. 26th {ICRC}, Vol.~1,
  Salt Lake City, 1999, p. 415.

\bibitem{bib:sibyll1}
R.~S. Fletcher, T.~Gaisser, P.~Lipari, T.~Stanev, Phys. Rev. D 50 (1994) 5710.

\bibitem{bib:sibyll}
R.~Engel, T.~Gaisser, P.~Lipari, T.~Stanev, Phys. Rev. D 46 (1992) 5013.

\bibitem{bib:geant}
S.~Agostinelli, et~al., Nucl. Instr. Meth. A 506 (2003) 250.

\end{thebibliography}

\end{document}